\documentclass[a4paper,11pt]{article}
\usepackage{pos}

\title{Loop Tree Duality with generalized propagator powers: numerical UV subtraction for two-loop Feynman integrals}
\ShortTitle{LTD with generalized propagator powers}

\author*[a]{Daniele Artico}

\affiliation[a]{Institut f\"ur Physik, Humboldt-Universit\"at zu Berlin, \\
Newtonstra\ss e 15, D-12489, Berlin, Germany}

\emailAdd{daniele.artico@physik.hu-berlin.de}

\abstract{An explicit Loop Tree Duality (LTD) formula for two-loop Feynman integrals with integer power of propagators is presented and used for a numerical UV divergence subtraction algorithm. This algorithm proceeds recursively and it is based on the $\mathcal{R}$ operator and the Hopf algebraic structure of UV divergences. After a short review of LTD and the numerical evaluation of multi-loop integrals, LTD is extended to two-loop integrals with generalized powers of propagators. The $\mathcal{R}$ operator and the tadpole UV subtraction are employed for the numerical calculation of two-loop UV divergent integrals, including the case of quadratic divergences.}

\FullConference{Loops and Legs in Quantum Field Theory (LL2024)\\
 14-19, April, 2024\\
Wittenberg, Germany\\}

\newcommand{\be}{\begin{equation}}
\newcommand{\ee}{\end{equation}}
\newcommand{\beq}{\begin{eqnarray}}
\newcommand{\eeq}{\end{eqnarray}}
\newcommand{\bea}{\begin{eqnarray}}
\newcommand{\eea}{\end{eqnarray}}

\usepackage{tikz-feynman}

\begin{document}
\maketitle


\section{Introduction: Loop Tree Duality}
\noindent
Loop Tree Duality (LTD) is a numerical integration method developed first for one-loop integrals \cite{Buchta:2015wna} and then for multiloop integrals \cite{Aguilera-Verdugo:2019kbz,Capatti:2019ypt,Capatti:2019edf}. It is based on the integration of the energy components of the loop momenta through the residue theorem. For multi-loop integrals, equivalent algorithms have been developed \cite{Aguilera-Verdugo:2019kbz, Capatti:2019ypt, Capatti:2019edf}.\\
Consider a $n$-loop integral 
\begin{equation}
\label{feynman integral}
I = \int \prod_{j = 1}^n \frac{d^4 k_j}{(2\pi)^4} \frac{N\left( \left\{k_j ,p_j,m_j\right\}\right)}{\prod_{i \in \textbf{e}}D_i\left( \left\{k_j ,p_j,m_j\right\}\right)}
\end{equation}
where $\left( \left\{k_j ,p_j,m_j\right\}\right)$ stands to indicate dependence on the loop momenta, the external momenta and the masses, $N\left( \left\{k_j ,p_j,m_j\right\}\right)$ is a polynomial in its variables, $\textbf{e}$ is the set of loop lines each having momentum $q_i$ and $D\left( \left\{k_j ,p_j,m_j\right\}\right)$ is the Feynman propagator (including the causal prescription) 
\begin{equation}
\label{Di}
D_i = (q_i^0)^2 - \vec{q_i}^2 - m_i^2 + i\delta.
\end{equation} 
The integral is taken to be in $D = 4$ as the goal is to perform numerical momentum-space integration, therefore all divergences are subtracted by suitable counterterms. The energy components are integrated one after the other in the order $k_1^0, k_2^0,..,k_n^0$ by closing the complex energy contours in the lower half-plane, thus multiplying each residue by $(-2\pi i)$. All residues corresponding to poles whose position in the lower/upper complex plane is not determined cancel pairwise \cite{Capatti:2019ypt,JesusAguilera-Verdugo:2020fsn}. For each loop momentum base \textbf{b} in the set of all possible bases $\mathfrak{B}$ made of propagator momenta , only a combination of residues survives in the final integrand, giving an overall prefactor $(-i)^n$. The integrand can be written as 
\begin{equation}
\label{ltd integrand}
I = (-i)^n \int \prod_{j = 1}^n \frac{d^3 k}{(2\pi)^3} \sum_{\textbf{b}\in \mathfrak{B}} \frac{N}{\prod_{i \in \textbf{b}}2E_i\prod_{i \in \textbf{e} \setminus \textbf{b}}D_i} \Bigg|_{\left\lbrace q_j^0 = \sigma_j^\textbf{b}E_j \right\rbrace_{j\in \textbf{b}}}
\end{equation}
where $\left\lbrace q_j^0 = \sigma_j^\textbf{b}E_j \right\rbrace_{j\in \textbf{b}}$ means that the energy componens of the loop momenta are determined by the residue corresponding to each summand. 
The LTD final result does not depend neither on the order of integration nor on the contour closure.\\
The cut propagators result in $E_i$ factors in the denominator of the summand. The other propagators become what are called the dual propagators, defined as
\begin{equation}
\label{dual propagator}
\frac{1}{D_i|_{\left\lbrace q_j^0 = \sigma_j^\textbf{b}E_j \right\rbrace_{j\in \textbf{b}}}} = \frac{1}{({q_i^0}_{\left\lbrace q_j^0 = \sigma_j^\textbf{b}E_j \right\rbrace_{j\in \textbf{b}}})^2 - (E_i)^2}.
\end{equation}
The dual propagators in Eq. \eqref{dual propagator} can become singular when the denominator is zero. The only surfaces in the integration space corresponding to physical thresholds have the form
\begin{equation}
\label{E-surface}
\eta_{\textbf{b},i} = p_{\textbf{b},i}^0 + E_i + \sum_{j \in \textbf{b}} \alpha_j E_j = 0.
\end{equation}
with $\alpha_j \in \left\lbrace 0,1 \right\rbrace$; while non physical thresholds cancel pairwise \cite{Aguilera-Verdugo:2019kbz,Capatti:2020ytd}. \\
For these singularities to appear, the energy component $ p_{\textbf{b},i}^0 $ should be negative and the squared modulus of the combination of external momenta for each surface has to be greater than the masses of the particles put on shell.
Given that the energies $E_i = \sqrt{\vec{q_i}^2 + m_i^2 - i\delta}$ have a negative imaginary part because of the $- i\delta$ in the square root, numerical evaluation of Feynman integrals with LTD determines in which direction the deformation of the integration contour in the complex plane has to point in order to have a correct value \cite{Capatti:2019edf}.\\ 

\section{Two-loop LTD for general powers of propagators}
\noindent
Residues of Feynman integrals with general propagator powers in the LTD framework have been a subject of study in Refs. \cite{JesusAguilera-Verdugo:2020fsn,
bierenbaum2012treeloop,aguileraverdugo2021stroll, Aguilera_Verdugo_2021,Aguilera_Verdugo_2020}. In this section, explicit formulas are presented for all two-loop tadpoles. The cancellation of nested residues is still valid, as conceptually raised propagator powers are limits of Feynman integrals with all unit powers when two propagators are the same. The arguments behind the formula for two-loop tadpoles are then used to write explicit LTD formulas which are valid for all two-loop integrals.
\subsection{Tadpoles}
Consider the two-loop tadpole 
\begin{equation}
\label{two-loop tadpole}
I_3^{a,b} = \int \frac{d^4 l_1 d^4 l_2}{(2\pi)^{8}} \frac{1}{(l_1^0-E_1)^a(l_1^0+E_1)^a(l_2^0-E_2)^b(l_2^0+E_2)^b(l_1^0+l_2^0-E_3)(l_1^0+l_2^0+E_3)}
\end{equation}
assumed to be UV and IR finite. \\
The LTD theorem produces three residues, while two others cancel. For $a = 2$ the integration yelds 
\begin{equation}
\label{2.1.1}
\small
I_{3,1,1}^{2,b} = - \int \frac{d^3 \vec{l_1} d^3 \vec{l_2}}{(2\pi)^{6}} \frac{1}{2E_3} \frac{1}{(b-1)!} \left.\left[ \partial_{l_2^0}^{b-1} \left( \frac{1}{(E_3-l_2^0-E_1)^2(E_3-l_2^0+E_1)^2(l_2^0+E_2)^b} \right) \right] \right\vert_{l_2^0\rightarrow E_2 },
\end{equation} 
\begin{equation}
\label{2.1.2}
I_{3,1,2}^{2,b} = - \int \frac{d^3 \vec{l_1} d^3 \vec{l_2}}{(2\pi)^{6}} \frac{1}{2E_3}  \left.\left[ \partial_{l_2^0} \left( \frac{1}{(E_3-l_2^0-E_1)^2(l_2^0-E_2)^b(l_2^0+E_2)^b} \right) \right] \right\vert_{l_2^0\rightarrow E_1+E_3}
\end{equation}
\begin{equation}
\label{2.2.1}
\small
I_{3,2,1}^{2,b} = - \int \frac{d^3 \vec{l_1} d^3 \vec{l_2}}{(2\pi)^{6}} \frac{1}{(b-1)!}  \left.\left[ \partial_{l_2^0}^{b-1}\partial_{l_1^0} \left( \frac{1}{(l_1^0+E_1)^2(l_2^0+E_2)^b(l_1^0+l_2^0+E_3)(l_1^0+l_2^0-E_3)} \right) \right] \right\vert_{l_2^0\rightarrow E_2}^{l_1^0\rightarrow E_1} 
\end{equation}
Given these premises, the elements of the LTD formula for tadpoles with generalized propagators and numerators are presented.  When all three propagator powers are taken to be general integer numbers, the terms $I^{2,b}_{3,1,1}$ in Eq. \eqref{2.1.1} and $I^{2,b}_{3,2,1}$ in Eq. \eqref{2.2.1} are slightly modified to get the correct integrands
\begin{equation}
\label{RP3P2}
\footnotesize
I_{3,1,1}^{a,b,c} = \frac{-1}{(b-1)!(c-1)!} \left.\left[ \partial_{l_2^0}^{b-1} \left( \partial_{l_1^0}^{c-1} \left( \left. \frac{N(l_i,m_i)}{(l_1^0-E_1)^a(l_1^0-E_1)^a(l_2^0+E_2)^b(l_1^0+l_2^0+E_3)^c} \right) \right\vert_{l_1^0\rightarrow E_3-l_2^0} \right) \right] \right\vert_{l_2^0\rightarrow E_2 }
\end{equation}
and
\begin{equation}
\label{RP1P2}
\footnotesize
I_{3,2,1}^{a,b,c} = \frac{-1}{(a-1)!(b-1)!} \left.\left[ \partial_{l_2^0}^{b-1} \left( \partial_{l_1^0}^{a-1} \left( \left. \frac{N(l_i,m_i)}{(l_1^0-E_1)^a(l_2^0+E_2)^b(l_1^0+l_2^0-E_3)^c(l_1^0+l_2^0+E_3)^c} \right) \right\vert_{l_1^0\rightarrow E_1} \right) \right] \right\vert_{l_2^0\rightarrow E_2 }.
\end{equation}
The other term is the analogous to $C^{2,b}_{3,1,2}$ in Eq. \eqref{2.1.2} and it is slightly different in its explicit formula. The reason is that it is obtained by taking the residue first in $l_1^0 = E_3-l_2^0$ and then in $l_2^0 = E_3+E_1$: the cut propagators are the first and the third one. After taking the first residue the power $a$ is modified to $a+c-1$: the integrand obtained when taking this into account is 
\begin{equation}
\label{RP3P1}
\small
\begin{split}
I_{3,2,1}^{a,b,c} = \frac{-1}{(a+c-2)!(c-1)!} \left.\left[ \partial_{l_2^0}^{a+c-2} \left( \partial_{l_1^0}^{c-1} \left( \left. \frac{N(l_i,m_i)}{(l_1^0-E_1)^a(l_1^0+E_1)^a(l_2^0+E_2)^b(l_2^0-E_2)^b} \right.\right.\right. \right. \right. \\ \left. \left.\left.\left.\left. \frac{1}{(l_1^0+l_2^0+E_3)^c} \right) \right\vert_{l_1^0\rightarrow E_3-l_2^0} (l_2^0-E_3-E_1)^{a+c-1}\right) \right] \right\vert_{l_2^0\rightarrow E_1+E_3 }.
\end{split}
\end{equation}
These three elements are the tree summands in the LTD formula of two-loop tadpoles with raised propagators. This result is checked against the result given by MATAD \cite{Steinhauser_2001} in the following table. The last row is calculated for $p = (1,0,0,0)$.
\begin{table}[!h]
\begin{center}
\begin{tabular}{|c|c|c|c|c|}
\hline 
Integral & N & $m_{UV}$ & LTD result & reference \\ 
\hline 
$I_{222}$ & $l_1 \cdot l_2$ & $1.73$ & $-4.4648(8) \cdot 10^{-6}$  & $-4.46629 \cdot 10^{-6}$  \\ 
\hline 
$I_{232}$ & $l_1 \cdot l_2$ & $1.73$ & $ 1.7108(1) \cdot 10^{-7}$  & $1.7107 \cdot 10^{-7}$  \\
\hline 
$I_{332}$ & $(l_1 \cdot l_1)(p\cdot l_1)(p \cdot l_2)$ & $1.73$ & $ 3.3668(1) \cdot 10^{-8}$  & $3.36672 \cdot 10^{-8}$  \\
\hline 
\end{tabular} 
\caption{LTD integration of two-loop tensor tadpoles checked against MATAD.}
\end{center}
\end{table}
\subsection{Adding external kinematics}
An $l$-loop integral has the topology of an $l$-loop tadpole, but to each line external legs can be attached. For the 2-loop case, only one tadpole topology exists. In presence of the external legs, each configuration above has more sub-configurations as to each propagator with the same loop momenta dependence but different combinations of external momenta correspond to a different residue. The formula for each of the three configurations is very similar to the one for each tadpole cut. Each propagator is written as
$$ \frac{1}{(l^0_i-p_i^0)^2-E_i^2}$$
where $l_i$ can be $l^0_1$, $l^0_2$ or $l^0_1+l^0_2$. In the next sectionst, the original loop-integrand is denoted with $\mathcal{F}$.
\subsubsection{Cut $l_1 - p_{i}$ and $l_2 - p_{j}$ }
The summand resulting from this cut choice corresponds to the one in Eq. \eqref{RP1P2}. Denote by $a$ and $b$ the exponents of the first and second cut propagators; then the 
residue of the two poles is
\begin{equation}
\label{cut12}
\small
R_{1ij} = \frac{-1}{(a-1)!(b-1)!} \left.\left[ \partial_{l_2^0}^{b-1} \left( \partial_{l_1^0}^{a-1} \left( \left. \mathcal{F}\cdot(l_1^0-p_i^0-E_i)(l_2^0-p_j^0-E_j) \right) \right\vert_{l_1^0\rightarrow E_i+p_i^0} \right) \right] \right\vert_{l_2^0\rightarrow E_j+p_j^0 }.
\end{equation}
\subsubsection{Cut $l_1 + l_2 - p_{i}$ and $l_2 - p_{j}$ }
This configuration corresponds to equation Eq. \eqref{RP3P2} and is no different in principle from the previous case. The resulting summand is
\begin{equation}
\label{cut32}
\small
R_{2ij} = \frac{-1}{(c-1)!(b-1)!} \left.\left[ \partial_{l_2^0}^{b-1} \left( \partial_{l_1^0}^{c-1} \left( \left. \mathcal{F}\cdot(l_1^0+l_2^0-p_i^0-E_i)(l_2^0-p_j^0-E_j) \right) \right\vert_{l_1^0\rightarrow E_i+p_i^0-l_2^0} \right) \right] \right\vert_{l_2^0\rightarrow E_j+p_j^0 }.
\end{equation}
where the exponents of the cut propagators are now $c$ and $b$.
\subsubsection{Cut $l_1 + l_2 - p_{i}$ and $l_1 - p_{j}$}
This configuration corresponds to Eq. \eqref{RP3P1} and it is different from the previous two for the same reason explained for the tadpole case (the differentiation when taking the residue in $l_1^0$ modifies the exponents of the propagators). The summand resulting from this cut is 
\begin{equation}
\label{cut31}
\small
\begin{split}
R_{3ij} = \frac{-1}{(a+c-2)!(c-1)!} \left.\left[ \partial_{l_2^0}^{a+c-2} \left( \partial_{l_1^0}^{c-1} \left( \left. \mathcal{F} \cdot (l_1^0+l_2^0-p_i^0-E_i)^c \right)\right\vert_{l_1^0\rightarrow E_i+p_i^0-l_2^0}\right. \right. \right.  \\ \cdot \left. \left.\left. (l_2^0-E_i-E_j-p_i^0+p_j^0)^{a+c-1}\right) \right] \right\vert_{l_2^0\rightarrow E_i+E_j+p_i^0-p_j^0 }
\end{split}
\end{equation}
where the exponents of the cut propagators are $c$ and $a$. By using the same rule of modifying derivatives when a $l_i$ dependence is introduced after $l_{i-1}^0$ integration, an algorithm for general loop number could be derived.\\
As an example of two-loop LTD application to integrals with raised exponents of propagators, the calculation of the following integral is presented:
\begin{equation}
\label{Lina's integral}
I = \int \frac{d^4 l_1 d^4l_2}{(2\pi)^8} \frac{1}{[l_1^2-m^2][(l_1+p_2)^2-m^2][(l_1+l_2)^2-m^2][l_2^2-m^2]{[(l_2-k)^2-m^2]^2}}
\end{equation}
where $p_2^2 = 0$ and $k^2 = 100$. The mass $m$ is the top quark mass, taken to be $1.73$.\\
There are 4 cuts belonging to the first configuration, two cuts belonging to the second and two cuts belonging to the third. In total, there are 8 summands.
The LTD numerical integration gives the result
\begin{equation}
I_{LTD} = (1.5114(1) \pm 1.3922(1)i)  10^{-7}
\end{equation}
while \textsc{pySecDec} gives
\begin{equation}
I_{SD} = (1.513(6) \pm 1.392(6)i)  10^{-7} .
\end{equation}

\section{The Bogoliubov $\mathcal{R}$ operator and tadpole subtraction}
\subsection{The $\mathcal{R}$ operator}
The Bogoliubov $\mathcal{R}$ operator is introduced following the derivation by Connes and Kreimer \cite{Kreimer1,Kreimer2,Kreimer3}. Given a graph $\Gamma$, the renormalized value of the log-divergent Feynman integral $U(\Gamma)$ associated to the diagram is defined recursively by
\begin{equation}
\label{R operator}
\mathcal{R}\left(\Gamma\right) = U(\Gamma) - \sum_{\gamma_i \in \Gamma} T(\gamma_i)\mathcal{R}\left(\Gamma/\gamma_i\right)
\end{equation}
where $\gamma_i$ is a divergent subgraph included in $\Gamma$, or $\Gamma$ itself. The operator $T$ takes the value of the Feynman integral in another kinematical point, e.g. a tadpole. In this case the counterterms are products of tadpoles with integrals with a lower loop number. This UV divergence subtraction is local and suitable for a momentum-space based numerical integration technique such as LTD.\\
The operator in Eq. \eqref{R operator} for for one- and two-loop 1PI Feynman integrals gives
\begin{equation}
\label{renormalized one loop}
\mathcal{R} \left( \Gamma^1 \right) = U\left( \Gamma^1 \right) - T \left( \Gamma^1 \right)
\end{equation}
and
\begin{equation}
\label{renormalized one loop}
\mathcal{R} \left( \Gamma^2 \right) = U\left( \Gamma^2 \right) - T \left( \gamma^1 \right) U\left( \Gamma^2/\gamma^1 \right)+ T \left( \gamma^1 \right) T\left( \Gamma^2/\gamma^1 \right) - T\left( \Gamma^2 \right)
\end{equation}
when the two-loop integral has also a one-loop subdivergence.\\
As an example, consider the UV divergence subtraction of the integral
\begin{equation}
\label{I(p^2,m^2)}
I(p^2,m^2) = \int \frac{d^D l_1 d^D l_2}{(2\pi)^{2D}} \frac{1}{[l_1^2-m^2][(l_1+p)^2-m^2][(l_1+l_2)^2-m^2][l_2^2-m^2]}
\end{equation}
which is UV divergent when both loop momenta are very large, and also when just $l_2$ grows to infinity. Following the subtraction strategy designed in the previous paragraphs, the integral 
\begin{equation}
\label{two loop example}
I_R(p^2,m^2,m_{UV}^2) = I(p^2,m^2) - C_1(p^2,m^2,m_{UV}^2) + C_2(m_{UV}^2) - C_3(m_{UV}^2)
\end{equation} 
is finite. The three counterterms are
\begin{equation}
\label{C_1}
C_1(p^2,m^2,m_{UV}^2) = \int \frac{d^D l_1 d^D l_2}{(2\pi)^{2D}} \frac{1}{[l_1^2-m^2][(l_1+p)^2-m^2][l_2^2-m_{UV}^2]^2}
\end{equation}
\begin{equation}
\label{C_2}
C_2(m_{UV}^2) = \int \frac{d^D l_1 d^D l_2}{(2\pi)^{2D}} \frac{1}{[l_1^2-m_{UV}^2]^2[l_2^2-m_{UV}^2]^2}
\end{equation}
\begin{equation}
\label{C_3}
C_3(m_{UV}^2) = \int \frac{d^D l_1 d^D l_2}{(2\pi)^{2D}} \frac{1}{[l_1^2-m_{UV}^2]^2[(l_1+l_2)^2-m_{UV}^2][l_2^2-m_{UV}^2]}.
\end{equation}
The integral in Eq.\eqref{two loop example} can be evaluated in $D = 4$ using LTD in the generalized propagator powers framework. With $p^2 = 100$, $m^2 = 1$ and $m_{UV}^2=226$ the result is
\begin{equation}
\label{two loop example LTD}
I_R(100,1.0,226.0)_{\textrm{LTD}} = (-1.9668(3) - 4.6483(2)i) · 10^{-4}
\end{equation} 
cross-checked with \textsc{PySecDec}
\begin{equation}
\label{two loop example pySD}
I_R(100,1.0,226.0)_{\textrm{SD}} = (-1.96767(1) - 4.64838(1)i) · 10^{-4}
\end{equation} 
\subsection{The subtraction of divergences using modified integrands}
The Hopf algebraic properties of UV scalar log-divergent integrals are well established and lead to the algorithmic production of local counterterms. The generalization to power divergent integrals can be obtained with an alternative derivation.
Consider the two-loop integrand
\begin{equation}
\label{2l power div}
\frac{d^Dk_1d^Dk_2N\left(k_1,k_2,p_i \right)}{P_1D\left(k_1,k_2,p_i,m_i \right)}
\end{equation}
where $P_1$ is a propagator. Supposing the loop integral is UV divergent when the loop momentum in $P_1$ goes to infinity, considering the modified propagator $\tilde{P_1}$ where the external momenta dependence of $P_1$ is removed and computing
\begin{equation}
\label{modified integrand subtraction}
\frac{d^Dk_1d^Dk_2N\left(k_1,k_2,p_i \right)}{P_1D\left(k_1,k_2,p_i,m_i \right)} - \frac{d^Dk_1d^Dk_2N\left(k_1,k_2,p_i \right)}{\tilde{P_1}D\left(k_1,k_2,p_i,m_i \right)} = \frac{d^Dk_1d^Dk_2N\left(k_1,k_2,p_i \right)(\tilde{P_1}-P_1)}{P_1\tilde{P_1}D\left(k_1,k_2,p_i,m_i \right)} 
\end{equation}
leads to an integral with degree of divergence for the loop momentum in $P_1$ going to infinity that is one unit lower than the original integrand. An iterated application of this strategy starting allows for the generation of local counterterms that make the integral finite and that have a simplified dependence on external momenta.\\
The advantage of this second method is that it is possible to generalize it to power divergent integrals.
\subsubsection{Power-divergences: the sunrise integral}
Consider the quadratic-divergent two-loop equal-mass sunrise integrand in four dimensions
\begin{equation}
\label{2Lsunrise}
I =  \frac{1}{[(k_1+p)^2-m^2][k_2^2-m^2] [(k_1+k_2)^2-m^2][(k_1)^2-m^2]^0} = \frac{1}{P_1P_2P_3\tilde{P_1}^0} = \frac{1}{P_1P_2P_3}
\end{equation}
where the modified propagator for the UV divergence subtraction $\tilde{P_1}$ has already been defined. The overall quadratic divergence when both loop momenta are large can be brought to a logarithmic one by subtracting the counterterm
\begin{equation}
\label{c1-sunrise}
I_{c_1} =   \frac{1}{\tilde{P_1}P_2P_3} - \frac{2k_1\cdot p + p^2}{\tilde{P_1}^2P_2P_3}.
\end{equation}
Indeed, power counting on 
\begin{equation}
\label{I-c1}
I-I_{c_1} =  \frac{(2k_1\cdot p + p^2)^2}{P_1\tilde{P_1}^2P_2P_3}
\end{equation}
proves that the overall UV divergence is now logarithmic. The remaining UV divergence can be removed by the $\mathcal{R}$ operator or modified integrand subtraction. The counterterm to subtract is 
\begin{equation}
\label{c2-sunrise}
I_{c_2} =    \frac{(2k_1\cdot p + p^2)^2}{P_1\tilde{P_1}^2P_2^2} -  \frac{(2k_1\cdot p + p^2)^2}{\tilde{P_1}^3P_2^2} +  \frac{(2k_1\cdot p + p^2)^2}{\tilde{P_1}^3P_2P_3}.
\end{equation}
The finite integral 
\begin{equation}
I_{\rm{fin}} = I-I_{c_1}-I_{c_2}
\end{equation}
can be computed with LTD thanks to local cancellation of UV divergences at the integrand level. For $p^2 = 400$ and $m^2 = 4$ the result is
\begin{equation}
I_{\rm{fin,LTD}} = (0.44934(3) + 0.018403(9)i)
\end{equation}
consistent with the \textsc{PySecDec} result
\begin{equation}
I_{\rm{fin,SD}} = (0.44943302 + 0.018397497)i)
\end{equation}
reported without errors as the presence of few sectors allows for a precision of 1 part per $10^{13}$. 

\section{Conclusion and outlook}
\noindent
In this conference proceeding, explicit formulas for LTD at two-loop in the presence of raised propagator powers are provided and tested against established numerical methods. These formulas are then used to provide a regularization algorithm both for log-divergent integrals and power-divergent ones. \\
Using the same approach of generalizing the (easier) tadpole formulas, the extension to higher loop number is natural. A different and interesting direction is the application of LTD techniques to linear propagators, such as the ones appearing for IR counterterms or in gravitational wave physics.
\section*{Acknowledgments}

\noindent DA is funded by the Deutsche Forschungsgemeinschaft (DFG, German Research Foundation) - Projektnummer 417533893/GRK2575 "Rethinking Quantum Field Theory". The author thanks Maria C. Sevilla, Dirk Kreimer, Lorenzo Magnea, Peter Uwer and Yingxuan Xu for the useful discussions.



\end{document}